\documentclass[times, twoside]{StyleArXiv}
\usepackage{booktabs}
\usepackage{multicol,multirow}
\usepackage{orcidlink}
\usepackage{subcaption}
\usepackage{cleveref}
\usepackage{graphicx}%
\usepackage{multirow}%
\usepackage{amsmath,amssymb,amsfonts}%

\leadauthor{Siddhad} 

\newcommand\numberthis{\addtocounter{equation}{1}\tag{\theequation}}

\begin{document}

\title{Modified TSception for Analyzing Driver Drowsiness and Mental Workload from EEG}
\shorttitle{Modified TSception for Analyzing Driver Drowsiness and Mental Workload from EEG}

\author[1,\Letter]{Gourav~Siddhad~\orcidlink{0000-0001-5883-3863}}
\author[2]{Anurag~Singh}
\author[3]{Rajkumar~Saini~\orcidlink{0000-0001-8532-0895}}
\author[4]{Partha~Pratim~Roy~\orcidlink{0000-0002-5735-5254}}
\affil[1]{Department of Computer Science and Engineering, Indian Institute of Technology, Roorkee, Uttarakhand, 247667, India}
\affil[2]{Department of Computer Science and Engineering, OP Jindal University, Raigarh, Chhattisgarh, 496109, India}
\affil[3]{Department of Computer Science, Electrical and Space Engineering, Luleå Tekniska Universitet, Luleå, 97187, Sweden}
\affil[4]{Department of Computer Science and Engineering, Indian Institute of Technology, Dhanbad, Jharkhand, 826007, India}

\maketitle



\begin{abstract}
Driver drowsiness is a leading cause of traffic accidents, necessitating real-time, reliable detection systems to ensure road safety. This study proposes a Modified TSception architecture for robust assessment of driver fatigue and mental workload using Electroencephalography (EEG). The model introduces a five-layer hierarchical temporal refinement strategy to capture multi-scale brain dynamics, surpassing the original TSception's three-layer approach. Key innovations include the use of Adaptive Average Pooling (ADP) for structural flexibility across varying EEG dimensions and a two-stage fusion mechanism to optimize spatiotemporal feature integration for improved stability. Evaluated on the SEED-VIG dataset, the Modified TSception achieves 83.46\% accuracy, comparable to the original model (83.15\%), but with a significantly reduced confidence interval (0.24 vs. 0.36), indicating better performance stability. The architecture's generalizability was further validated on the STEW mental workload dataset, achieving state-of-the-art accuracies of 95.93\% and 95.35\% for 2-class and 3-class classification, respectively. These results show that the proposed modifications improve consistency and cross-task generalizability, making the model a reliable framework for EEG-based safety monitoring.
\end{abstract}


\begin{corrauthor}
g\_siddhad\at cs.iitr.ac.in
\end{corrauthor}


\section{Introduction}
\label{sec_intro}

Electroencephalography (EEG) is a non-invasive technique for recording the brain's electrical activity with significant applications across neuroscience, clinical diagnosis, and Brain-Computer Interfaces (BCIs). By capturing signals across distinct frequency bands---delta, theta, alpha, beta, and gamma -- via scalp electrodes, EEG provides the high temporal resolution necessary to monitor rapid neural dynamics in real-time. Compared to other neuroimaging modalities, EEG is highly valued for its portability and cost-effectiveness, making it ideal for assessing cognitive states such as driver drowsiness and mental workload.

In clinical practice, EEG serves as a cornerstone for diagnosing neurological conditions, including epilepsy, alzheimer's disease, and various sleep disorders~\cite{amer2023eeg}. Its capacity for real-time monitoring is particularly critical for conditions requiring precise temporal resolution~\cite{michel2019eeg}. Beyond clinical use, EEG is instrumental in BCIs~\cite{craik2019deep}, neuromarketing, and psychology, where Event-Related Potentials (ERPs) are used to characterize stimulus-driven brain processes. Furthermore, EEG provides an objective measure of cognitive load, making it a vital tool for assessing mental workload in high-demand environments.

Driver drowsiness is a global contributor to road accidents, and its timely detection is pivotal for improving road safety. Unlike fatigue, which is a broader state, drowsiness represents a specific transition toward sleep that EEG can detect directly through brain activity. This allows for the identification of early neural markers, such as alpha spindles and theta bursts, before external physical signs become visible to behavioral or visual monitoring systems.

Effective analysis of these complex signals requires sophisticated processing, including artifact removal, feature extraction, and classification. Recently, Deep Learning (DL) has emerged as a superior approach to traditional Machine Learning (ML) for EEG classification. Architectures such as Convolutional Neural Networks (CNNs) and Recurrent Neural Networks (RNNs) have demonstrated success in emotion recognition and seizure detection~\cite{craik2019deep}. A primary advantage of DL is its ability to perform automated feature extraction directly from raw EEG data~\cite{roy2019deep}. In the context of drowsiness, CNNs are particularly effective as they can learn biologically meaningful patterns that serve as reliable indicators of a subject's state~\cite{cui2022compact}.

Despite this progress, challenges in EEG-based detection remain, particularly regarding inter-individual variability. The inherent differences in brain activity between subjects complicate the development of calibration-free systems. Current research aims to improve generalizability, minimize sensor requirements, and explore multi-modal integration with other physiological measures~\cite{stancin2021review}.

While EEG offers a direct and objective measure of brain activity, a critical gap remains: current DL models often exhibit high performance instability across subjects. Furthermore, the lack of a unified architecture capable of generalizing across different cognitive tasks, such as drowsiness and mental workload, limits real-world implementation. Consequently, this research aims to bridge these gaps by proposing a Modified TSception architecture. By enhancing temporal and spatial feature extraction, this study develops a robust system evaluated on the SEED-VIG (drowsiness) and STEW (mental workload) datasets.

To address the limitations inherent in existing methodologies, three primary research objectives are pursued in this study, with a central focus on enhancing the reliability and versatility of EEG-based classification. First, stability enhancement is sought through the development of a robust architecture designed to minimize performance inconsistency and yield more reliable confidence intervals in comparison to current models. Second, cross-task generalizability is evaluated to determine the effectiveness of a modified spatiotemporal network in detecting both driver drowsiness and fluctuating levels of mental workload. Finally, architectural optimization is addressed by investigating the role of hierarchical temporal layers and adaptive pooling mechanisms in capturing the complex, non-linear patterns characteristic of multi-channel EEG signals. The contributions of this paper are as follows: 
\begin{itemize} 
    \item Introduction of a Modified TSception architecture, which incorporates additional temporal layers for richer hierarchical feature extraction.
    \item Strategic implementation of Adaptive Average Pooling (ADP) within temporal and spatial pathways to maintain fixed output dimensions and mitigate overfitting.
    \item Integration of a two-stage fusion mechanism designed to optimize the synergy between extracted spatiotemporal features.
    \item Cross-domain evaluation of the proposed model using the SEED-VIG and STEW datasets to verify performance across different EEG-based tasks.
    \item Validation of performance stability, demonstrating that the proposed model achieves accuracy comparable to the original TSception while significantly reducing the Confidence Interval (CI) for real-world reliability.
\end{itemize}

The remainder of this paper is structured as follows: Section~\ref{sec_background} reviews the literature on EEG-based drowsiness detection. Section~\ref{sec_method} details the Modified TSception architecture and its design rationale. Section~\ref{sec_results} presents the experimental results and comparative evaluation, and Section~\ref{sec_conclusion} concludes the paper with directions for future work.


\section{Background}
\label{sec_background}

This section reviews the existing literature on EEG-based drowsiness detection and mental workload assessment. The review encompasses fundamental signal processing concepts, traditional ML methodologies, and modern DL architectures, while also addressing critical challenges such as inter-subject variability and the development of portable, low-cost monitoring devices.

Drowsiness detection remains a vital research area for enhancing road safety, as EEG signals provide a direct, objective measure of neurological transitions in alertness. Traditional ML approaches in this domain rely heavily on manual feature engineering. For instance, Tarafder et al.~\cite{tarafder2022drowsiness} utilized ocular indices derived from eye movements to classify drowsy and alert states with 91.10\% accuracy using an ensemble-boosted trees model. Other researchers have integrated EEG with near-infrared spectroscopy (NIRS) to identify significant shifts in oxy-hemoglobin concentration and beta-band power during the onset of fatigue~\cite{nguyen2017utilization}. Furthermore, the combination of wavelet-based nonlinear features with eyelid movement data has demonstrated improved detection speed and accuracy~\cite{chen2015automatic}.

To address inter-individual variability, studies have explored cross-subject systems and channel selection. Cui et al.~\cite{cui2022compact} proposed a compact CNN to identify shared EEG features across subjects, achieving 73.22\% accuracy and revealing biologically explainable neural markers. Statistical channel selection has also been employed to optimize single-channel EEG systems, with a prefrontal device achieving 72.7\% accuracy in practical settings~\cite{ogino2018portable, balam2021statistical}.

DL has significantly advanced the field by enabling the automated learning of complex patterns directly from raw data~\cite{siddhad2024enhancing}. CNNs are currently the most prevalent architecture in EEG classification due to their ability to extract both spatial and temporal features. Studies using keras-based CNNs have achieved accuracies as high as 90.42\%~\cite{chaabene2021convolutional}, while more compact models have highlighted the capacity of CNNs to discover shared features related to drowsiness across diverse subject groups~\cite{cui2022compact}.

Hybrid models and specialized architectures have further pushed performance boundaries. Integrating EEG spectrograms with deep feature extraction has yielded 94.31\% accuracy~\cite{budak2019effective}, while deep rhythm-based approaches using EEG images and Long Short-Term Memory (LSTM) networks reached 97.92\%~\cite{turkoglu2021deep}. These results demonstrate the potential of integrating time-frequency transformations with sequential DL models.

Beyond CNNs, RNNs, particularly LSTMs, are well-suited for capturing the temporal dependencies inherent in sequential EEG data~\cite{walther2023systematic}. Although effective for monitoring transitions between alert and drowsy states, RNNs can be computationally expensive; however, the addition of attention mechanisms has been shown to improve classification efficiency~\cite{walther2023systematic}. Recently, Transformers have been integrated into hybrid architectures, such as CNN-Former~\cite{ding2024novel}, to leverage global dependencies alongside local feature extraction. Additionally, spatiotemporal image encoding—such as recurrence plots or Gramian Angular Fields (GAF)—has shown promise in enhancing cross-subject detection using CNNs~\cite{paulo2021cross}.

While research on mental workload is less extensive than drowsiness detection, it is a rapidly growing field. Workload classification often leverages similar DL architectures, including CNNs and hybrid models, alongside spectral power analysis and wavelet transforms. The STEW dataset serves as a primary resource for evaluating these models across varying levels of cognitive demand.

Existing methodologies face several limitations, including signal loss~\cite{samiee2014data}, the intrusiveness of high-density EEG caps~\cite{perkins2022challenges}, and a reliance on limited simulation data or subjective measures. To improve practicality, researchers have investigated non-hair-bearing (NHB) EEG areas, which show comparable accuracy to whole-scalp montages~\cite{wei2018subject}. The development of low-cost consumer headsets has also been a focus of systemic reviews, highlighting their potential for widespread adoption despite higher noise levels~\cite{larocco2020systemic}. EEG signals are inherently noisy and exhibit significant inter- and intra-subject variability. Techniques such as hierarchical clustering, subject-transfer frameworks, and data augmentation are frequently employed to mitigate these issues and reduce overfitting~\cite{wei2018subject, chinara2021automatic}.

Recent advancements in driver drowsiness detection have shifted from traditional ML (SVM, KNN) to DL models like EEGNet and the original TSception. While these models excel at capturing complex patterns, they often exhibit high performance volatility. For instance, a model may achieve 90\% accuracy on one subject but drop to 70\% on another, leading to wide CIs that make them unreliable for real-world safety deployment. Furthermore, many current methodologies are 'task-specific', failing to generalize when the cognitive demand shifts from fatigue to high mental workload.

In conclusion, while CNNs and hybrid architectures have achieved high accuracy, the challenge of cross-subject variability and performance consistency remains. There is a critical need for robust, generalizable models that can accurately assess both drowsiness and mental workload across diverse individuals and settings.


\section{Methodology}
\label{sec_method}

In this section, the methodologies employed in the study are detailed, with a specific focus on the architectural enhancements made to the original TSception model~\cite{ding2022tsception}. Both the foundational TSception architecture and the proposed modifications are presented, while the technical rationale for these changes is highlighted within the context of EEG-based drowsiness and mental workload detection.


\subsection{TSception Preliminaries}

TSception~\cite{ding2022tsception} is a DL architecture designed to capture both temporal and spatial features of EEG signals. It utilizes parallel convolutional layers to extract features across multiple temporal resolutions (temporal convolutions) and spatial configurations (spatial convolutions), crucial for analyzing EEG signals containing information across different time scales and electrode channels.

The architecture comprises key components: temporal convolution layers (Tception layers) employ three parallel convolutional blocks with varying kernel sizes, corresponding to window lengths of 0.5, 0.25, and 0.125 seconds (based on a 200 Hz sampling rate), to capture EEG signal dynamics at different time scales. Each block applies average pooling (pool size 8) for dimensionality reduction and uses Leaky ReLU as the activation function for efficient gradient propagation. Following temporal feature extraction, spatial convolutional blocks (Sception layers) capture interactions between EEG channels using different kernel sizes to analyze all subsets of channels. Batch normalization is applied after each spatial convolution. A fusion convolutional layer integrates the spatial and temporal features, with pooling and batch normalization. The final stage includes a fully connected layer with ReLU activation, dropout, and a final dense layer mapping features to the target classes (drowsy and awake).

\begin{figure*}[t!]
    \centering
    \begin{subfigure}[t]{0.2\textwidth}
        \centering
        \includegraphics[width=\linewidth]{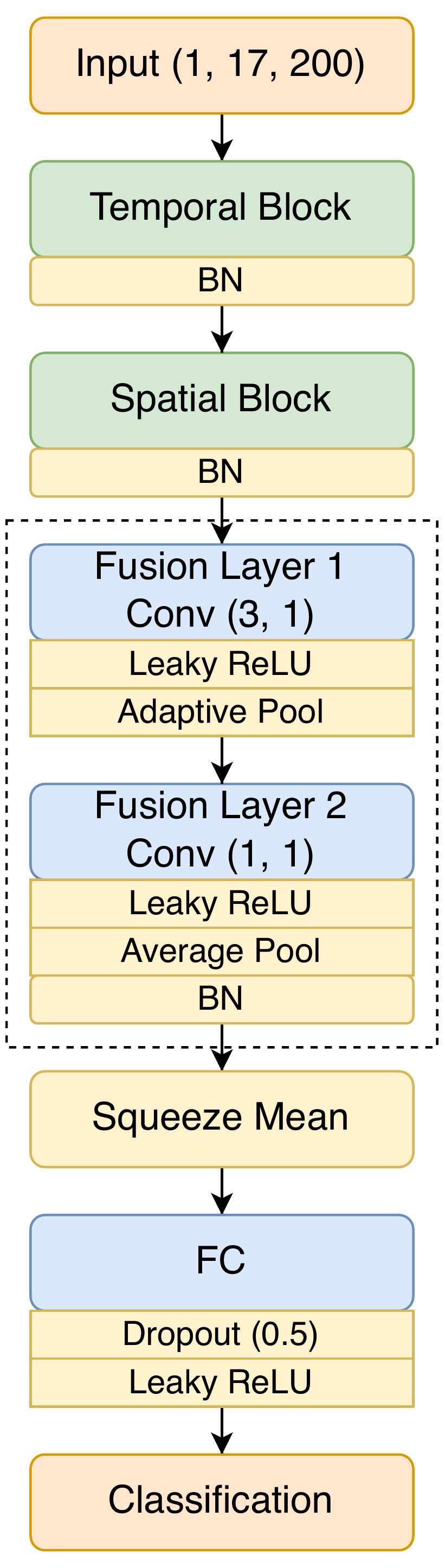}
        \label{fig_modTS_arch}
        \caption{}
    \end{subfigure}
    ~
    \begin{subfigure}[t]{0.4\textwidth}
        \centering
        \includegraphics[width=\linewidth]{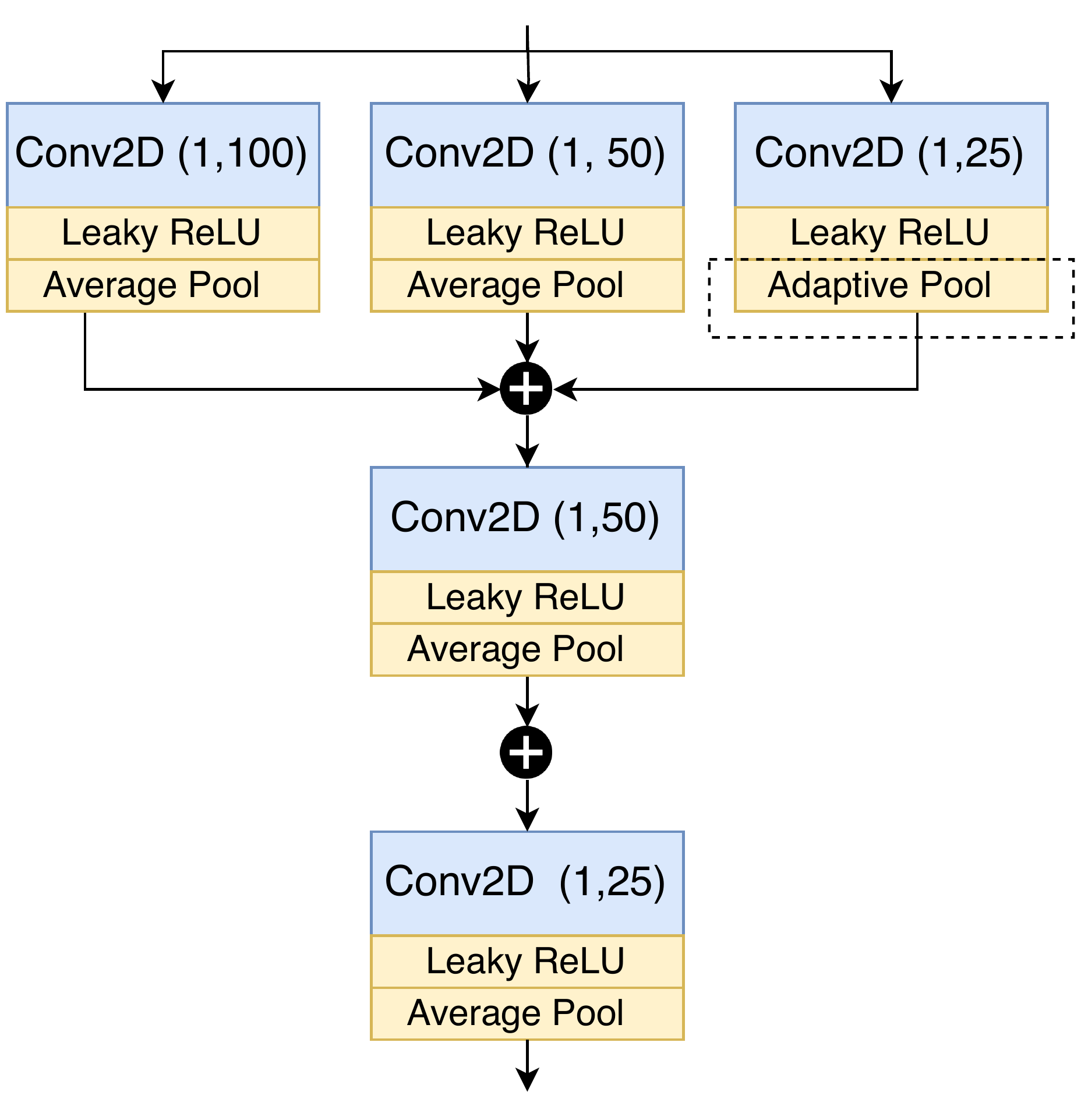}
        \label{fig_modTS_temporal}
        \caption{}
    \end{subfigure}
    ~
    \begin{subfigure}[t]{0.3\textwidth}
        \centering
        \includegraphics[width=\linewidth]{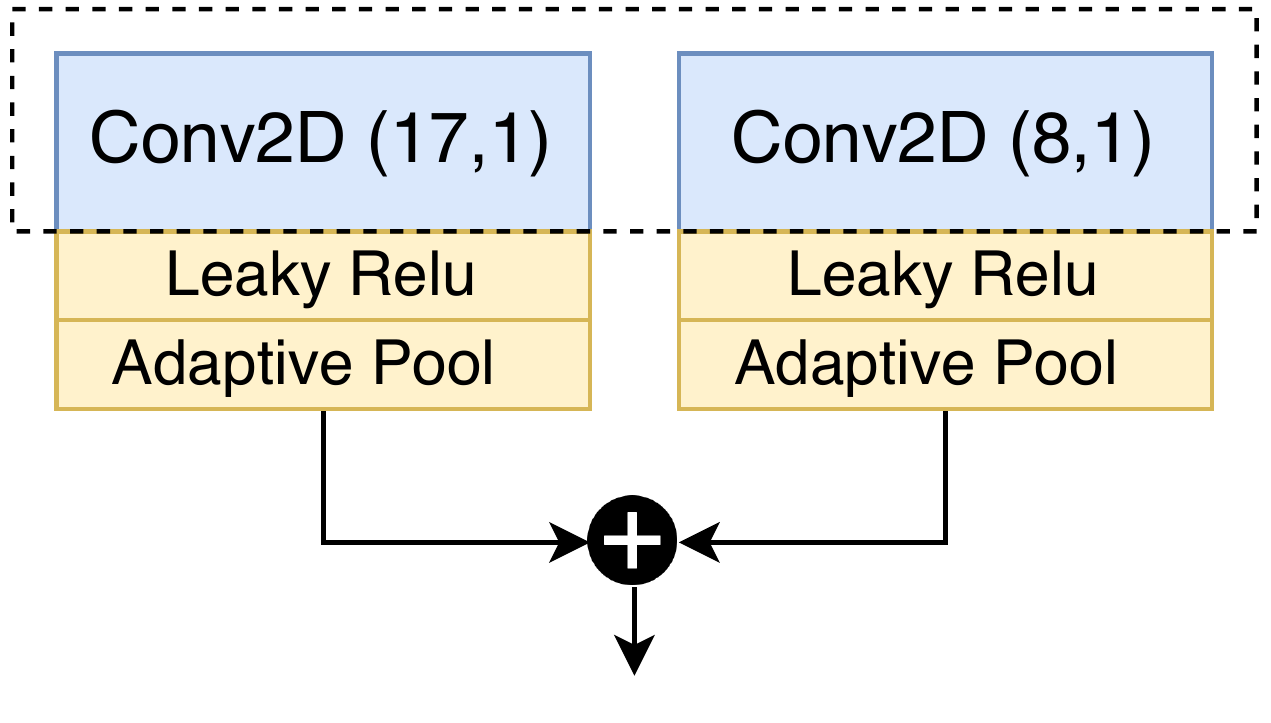}
        \label{fig_modTS_spatial}
        \caption{}
    \end{subfigure}
    \label{fig_modTS}
    \caption{Architectural details of the proposed Modified TSception model: (a) Overall architecture, (b) Temporal Block (Tception layer), and (c) Spatial Block (Sception layer). Dashed boxes represent the specific modifications made to the original TSception framework. Abbreviations: BN (Batch Normalization), FC (Fully Connected), Conv (Convolution), AP (Average Pooling), ADP (Adaptive Average Pooling), GAP (Global Average Pooling)}
\end{figure*}


\subsection{Modified TSception}

The Modified TSception model enhances the foundational architecture to provide more robust feature extraction across diverse datasets. The temporal pathway is expanded to five Tception layers. While the first three layers capture short, mid, and long-term rhythmic dynamics, the additional two layers use reduced kernel sizes to refine the feature maps. Spatial extraction is similarly improved through two Sception layers with variable-sized kernels to process full and partial channel configurations.

The model processes input EEG tensors of dimensions (1,17,200), representing a single-channel input from 17 electrodes over 200 samples. Hyperparameters include 15 filters for spatiotemporal layers and an FC network with 64 hidden units and a 0.5 dropout rate. The network is trained using cross-entropy loss with label smoothing. Three critical modifications drive the improved performance of the proposed model:

\textbf{Hierarchical Temporal Layers:} The use of five Tception layers instead of three allows the model to perform multi-scale temporal filtering. The first three layers capture standard rhythmic activity, while the additional two layers with halved sampling rates act as refinement filters to capture transient EEG bursts, such as alpha spindles, which are key indicators of the onset of drowsiness.

\textbf{Adaptive Average Pooling:} Unlike fixed pooling, which can be sensitive to the exact length of the input window, ADP ensures that the feature maps are down-sampled to a consistent size regardless of the input's sampling rate or channel count. This makes the architecture `device-agnostic' and contributes to the high generalizability seen in the STEW dataset results. The implementation of ADP in temporal layer 3 serves as a structural bottleneck that standardizes feature dimensionality. While the initial layers focus on broad temporal correlations, the third layer transitions to finer-scale analysis. ADP ensures that multi-scale features are down-sampled into a uniform representation, facilitating stable input for the two-stage fusion mechanism and addressing variance in temporal resolution across different recording sessions.

\textbf{Two-Stage Fusion:} A single fusion layer often fails to capture the complex relationship between where a signal originates (spatial) and how it changes over time (temporal). Proposed two-stage approach first integrates spatial features and then refines them using $1 \times 1$ pointwise convolutions to enhance inter-channel relationships, which directly led to the improved stability ($\pm 0.24$ CI) reported in Table~\ref{tab_results}.


\subsection{Layer Architecture of Modified TSception}

This subsection provides the mathematical formalization of the core components.
 
\textbf{Temporal Layer:}
The temporal layer extracts time-domain features using input $x$ and 2D convolution with kernel sizes $(1, k_t)$. $k_t$ is calculated as:
\begin{equation}
    k_t = \lfloor \text{inception window}[i] \times \text{sampling rate} \rfloor
\end{equation}
for the first three temporal layers, and for the remaining two temporal layers, the sampling rate halved.

The inception window is \([0.5, 0.25, 0.125]\). The extracted feature is activated by LeakyReLU, and down-sampled using AP, except for layer 3 which uses ADP. The output is concatenated and batch normalized (${BN}_t$). The temporal layer output is:
\begin{align*}
    z_\text{LReLU} &= \sigma({\text{conv2D}}(x, (1, k_t))) \\
    z_\text{pool} &= \text{AP}(z_\text{LReLU}) \quad \text{or} \quad \text{ADP}(z_\text{LReLU}) \\
    z_\text{BN} &= \text{BN}(z_\text{pool}) \\
    y_t &= z_\text{BN} \numberthis
\end{align*}

\textbf{Spatial Layer:}
The spatial layer captures spatial dependencies using 2D convolution with kernel sizes $(17, 1)$ and $(8, 1)$ for the two spatial layers. The extracted feature is activated by LeakyReLU and down-sampled using ADP. The output is concatenated and batch normalized ($\text{bn\_s}$). The spatial layer output is:
\begin{align*}
    z_{\text{LReLU}} &= \sigma({\text{conv2D}}(x, (1, k_t))) \\
    z_{\text{ADP}} &= \text{ADP}(z_{LReLU}) \\
    y_s &= \text{BN}(z_{\text{ADP}}) \numberthis
\end{align*}

\textbf{Fusion Layer:}
The fusion layer fuses temporal and spatial information. A 2D convolutional layer (kernel size $(3, 1)$) fuses information along the spatial dimension. After LeakyReLU, ADP/AP, and batch normalization ($\text{bn\_f}$), a Global Average Pooling (GAP) layer mitigates overfitting. A smaller fusion layer with a 1D convolutional layer (kernel size $(1, 1)$) enhances inter-channel relationships. The fusion layer outputs $y_{f1}$ and $y_{f2}$, are:

\begin{align*}
    z_{\text{LReLU}} &= \sigma({\text{conv2D}}(x, (3, 1))) \\
    z_{\text{ADP}} &= \text{ADP}(z_{\text{LReLU}}) \\
    y_{f1} &= \text{GAP}(\text{BN}(z_{\text{ADP}})) \numberthis
\end{align*}
\begin{align*}
    z_{\text{LReLU}} &= \sigma(\text{conv1D}(x, (1, 1))) \\
    z_{\text{AP}} &= \text{AP}(z_{\text{LReLU}}) \\
    y_{f2} &= \text{GAP}(\text{BN}(z_{\text{AP}})) \numberthis
\end{align*}
$y_{f2}$ is fed into fully connected layers with a softmax output.


\subsection{Comparison of Architectural Enhancements}

While both TSception and Modified TSception leverage parallel spatiotemporal pathways, the proposed model introduces significant refinements. By expanding to five Tception layers and replacing fixed pooling with ADP, the model achieves better scale invariance. The introduction of the $1 \times 1$ pointwise fusion layer and a deeper FC network (two hidden layers with dropout) further enhances the model's ability to extract stable, high-level features, directly improving generalizability across diverse cognitive tasks. Compared to established architectures like EEGNet and the original TSception, the Modified TSception offers three distinct advantages:

\textbf{Scale Invariance:} While EEGNet utilizes fixed-size kernels that may fail to capture transient bursts across different sampling rates, our five-layer hierarchy allows for simultaneous analysis of both sustained rhythmic activity and rapid, non-stationary EEG spindles.

\textbf{Hardware Agnosticism:} The transition from fixed AP to ADP removes the dependency on specific input dimensions. This allows the model to be deployed across different EEG headsets (e.g., research-grade vs. consumer-grade) without redesigning the network architecture.

\textbf{Enhanced Stability:} The two-stage fusion mechanism specifically addresses the high inter-subject variability found in traditional CNNs. By refining spatial features with pointwise convolutions before final classification, the model achieves a more stable feature representation, as evidenced by the significant reduction in performance variance across different subjects.


\section{Results and Discussion}
\label{sec_results}

This section presents the experimental findings of this study, evaluating the proposed Modified TSception architecture for driver drowsiness detection using the SEED-VIG dataset and mental workload assessment using the STEW dataset. Performance is quantified through accuracy and 95\% CIs, compared against traditional ML baselines and state-of-the-art DL models. The analysis explores the effectiveness of the proposed modifications, the resulting stability improvements, and the broader implications for reliable EEG-based monitoring.


\subsection{Experimental Data}
This study investigates the automated detection of driver drowsiness and mental workload using EEG. Two distinct datasets, SEED-VIG and STEW, are leveraged to achieve this goal. The generalizability of the proposed methodology across different cognitive states and experimental paradigms is evaluated through analysis of these datasets.

\textbf{SEED-VIG (Driver Drowsiness Dataset):} The SEED-VIG dataset~\cite{zheng2017multimodal} is an open-source resource for investigating driver vigilance. It comprises EEG recordings from 23 participants (12 males and 11 females) with a mean age of approximately 23.3 years (SD: 1.4 years) engaged in a simulated driving task designed to mimic real-world conditions. Data were acquired using a 17-channel montage focusing on temporal (FT7, FT8, T7, T8, TP7, TP8) and posterior (CP1, CP2, P1, PZ, P2, PO3, POZ, PO4, O1, OZ, O2) regions at a 1000~Hz resolution. Drowsiness was quantified via the PERCLOS (percentage of eyelid closure) metric, with a threshold of 0.5 defining ``awake'' versus ``drowsy'' states. To isolate cognitive features and remove physiological noise, signals were band-pass filtered between 1 and 75 Hz using a zero-phase Butterworth filter. Data were down-sampled from 1000 to 200 Hz to streamline computation while maintaining sufficient resolution for temporal convolution. Continuous data were segmented into one-second non-overlapping epochs, resulting in a standardized input tensor of (1, 17, 200). The resulting 40,710 samples were split into training (70\%), validation (15\%), and test (15\%) sets.

\textbf{STEW (Mental Workload Dataset):} To validate cross-task efficacy, the STEW dataset~\cite{lim2018stew} was utilized, which focuses on multitasking workload. Raw EEG data from 48 subjects (aged 18–40) were recorded during the simultaneous capacity (SIMKAP) test. Recordings were performed using an Emotiv EPOC headset (14 channels, 128~Hz sampling rate). The protocol included a 2.5-minute resting state (``No task'') and a 2.5-minute workload condition. Preprocessing for the STEW dataset involved several stages to mitigate the higher noise levels associated with consumer-grade headsets. After initial band-pass filtering, Independent Component Analysis (ICA) was performed to decompose the signal into statistically independent components, allowing for the identification and rejection of components corresponding to eye blinks, lateral eye movements, and temporal muscle activity. Furthermore, epochs were manually inspected and removed as `bad epochs' before the data were min-max scaled and segmented with a 0.5-second overlap to enhance the training set size. The final data tensor dimensions were (1, 14, 128) across 26,910 samples.


\subsection{Implementation Details}
Experiments were conducted on a DELL Precision 7820 Tower Workstation (Intel Xeon Silver 4216 CPU, NVIDIA RTX A4000 12GB GPU) running Ubuntu 22.04. Models were implemented in Python 3.12 using the PyTorch library. The Adam optimizer was employed ($\eta$ = 0.001, $\beta_1$ = 0.9, $\beta_2$ = 0.999). Key libraries included MNE-Python for EEG signal processing, scikit-learn~\cite{sklearn} for baseline SVM classification and metrics, and NumPy/Pandas for data manipulation. Training utilized 16-bit floating-point precision for model weight updates. Both EEGNet and the TSception variants were trained for 100 epochs with a batch size of 16 and a learning rate of $1e-4$. Stratified five-fold cross-validation was used to ensure robust performance evaluation.


\subsection{Classifiers}
To benchmark the proposed model, a balanced suite of classifiers was employed, integrating both traditional ML and modern deep learning architectures. A SVM was implemented using a Radial Basis Function (RBF) kernel to facilitate non-linear classification through high-dimensional input mapping~\cite{cortes1995support}. For neural network-based approaches, EEGNet was utilized as a compact CNN architecture that leverages depth-wise and separable convolutions for the efficient extraction of frequency-specific spatial filters~\cite{lawhern2018eegnet}.

The TSception model was incorporated as an original spatiotemporal architecture, specifically designed to learn emotional asymmetry and temporal representations via varying kernel sizes~\cite{ding2022tsception}. To assess enhanced feature learning efficiency, ConvNext was employed, representing a state-of-the-art CNN that adopts design principles from Transformers~\cite{liu2022convnet}. Furthermore, LMDA-Net, a lightweight multi-modal spatiotemporal model, was used for specialized EEG-based classification~\cite{miao2023lmda}. Finally, an attention-based Transformer network~\cite{siddhad2024efficacy} was included to evaluate the efficacy of global dependency modeling within the context of mental workload assessment.


\subsection{Performance Evaluation}

\begin{table}[!t]
    \centering
    \caption{Results of different methods on SEED-VIG for driver drowsiness detection and STEW for mental workload assessment with 95\% CI}
    \label{tab_results}
    \resizebox{\columnwidth}{!}{
    \begin{tabular}{lccc}
        \toprule
        \multirow{2}{*}{\textbf{Method}} & \textbf{SEED-ViG} & \multicolumn{2}{c}{\textbf{STEW}} \\
        & \textbf{2 Class} & \textbf{2 Class} & \textbf{3 Class} \\
        \midrule
        \textbf{SVM}~\cite{cortes1995support} & $65.52 \pm 0.02$ & $83.34 \pm 0.39$ & $83.21 \pm 0.28$ \\
        \textbf{Transformer}~\cite{siddhad2024efficacy} & $68.95 \pm 0.21$ & $95.32 \pm 0.00$ & $88.72 \pm 0.00$\\ 
        \textbf{EEGNet}~\cite{lawhern2018eegnet} & $80.74 \pm 0.75$ & $84.33 \pm 1.03$ & $83.64 \pm 1.33$\\
        \textbf{LMDA-Net}~\cite{miao2023lmda} & $81.06 \pm 0.99$ & $89.54 \pm 1.30$ & $ 90.41 \pm 1.05$\\
        \textbf{ConvNeXt}~\cite{liu2022convnet} & $81.95 \pm 0.61$ & $95.76 \pm 0.51$ & $95.11 \pm 0.80$\\
        \textbf{TSception}~\cite{ding2022tsception} & $83.15 \pm 0.36$ & $95.21 \pm 0.53$ & $94.73 \pm 0.30$\\
        \midrule
        \textbf{Modified TSception} & $\mathbf{83.46 \pm 0.24}$ & $\mathbf{95.93 \pm 0.54}$ & $\mathbf{95.35 \pm 0.55}$\\
        \bottomrule
    \end{tabular}}
\end{table}

\begin{figure*}[!t]
    \centering
    \includegraphics[width=\textwidth]{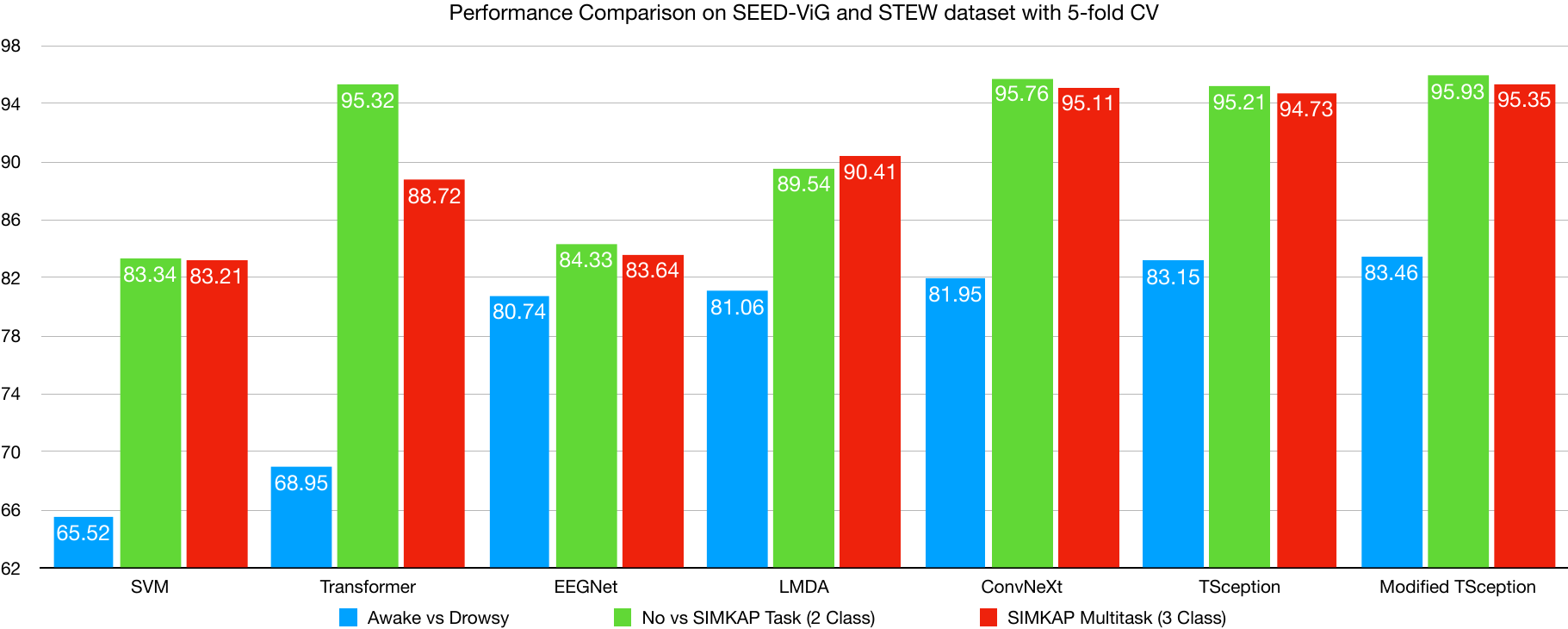}
    \caption{Barchart showing the performance comparison of different methods on the SEED-VIG for driver drowsiness detection and STEW dataset for mental workload assessment using 5-fold cross-validation.}
    \label{fig_chart}
\end{figure*}

Table~\ref{tab_results} and the corresponding bar chart (Fig.~\ref{fig_chart}) present a comparative performance analysis across the SEED-VIG and STEW datasets. Accuracy is reported alongside 95\% CIs to ensure a robust comparison. In a practical setting, a narrower CI implies that the system is more likely to maintain its advertised performance level regardless of the specific user's brainwave patterns, thereby reducing the risk of 'false negatives' in subjects who deviate from the population average.


\subsubsection{Driver Drowsiness Detection}

For binary classification on the SEED-VIG dataset, DL models significantly outperformed traditional ML. The Support Vector Machine (SVM)~\cite{cortes1995support} achieved the lowest accuracy (65.52\% $\pm$0.02). The Transformer model~\cite{siddhad2024efficacy} offered only a modest improvement to 68.95\% ($\pm$0.21), illustrating the limitations of general-purpose architectures for raw EEG analysis. In contrast, specialized DL models showed substantial gains: EEGNet~\cite{lawhern2018eegnet} achieved 80.74\% ($\pm$0.75), while LMDA-Net~\cite{miao2023lmda} and ConvNeXt~\cite{liu2022convnet} reached 81.06\% and 81.95\%, respectively. The original TSception model~\cite{ding2022tsception} provided the highest baseline accuracy at 83.15\% ($\pm$0.36), underscoring the effectiveness of its multi-scale feature extraction. Proposed Modified TSception achieved 83.46\% accuracy with a reduced CI of $\pm$0.24. While the accuracy gain is incremental (0.31\%), the 33.3\% reduction in the CI signifies improved performance stability and robustness, which are critical for safety-oriented real-world applications.


\subsubsection{Mental Workload Assessment}

On the STEW dataset, the Modified TSception was evaluated for both 2-class (resting vs. workload) and 3-class (low, medium, high) tasks. For the 2-class task, proposed model achieved a state-of-the-art accuracy of 95.93\% ($\pm$ 0.54\%), marginally surpassing ConvNeXt (95.76\%) and the original TSception (95.21\%). This trend continued in the 3-class task, where Modified TSception reached 95.35\% ($\pm$ 0.55\%). Interestingly, the Transformer model showed excellent results for binary workload classification (95.32\%), suggesting that the structured temporal dynamics of cognitive load are well-suited to attention-based sequence modeling. While the SVM performed reasonably well on STEW (83.34\%) compared to SEED-VIG, it remained significantly behind the DL-based approaches.


\subsection{Overall Performance and Discussion}

The Modified TSception model demonstrates superior performance across both datasets, highlighting its effectiveness for unified cognitive state monitoring. A progressive improvement in accuracy is observed as the methodology transitions from traditional ML methods to specialized DL architectures tailored for EEG dynamics. It is observed that DL models consistently outperform the SVM, particularly on the SEED-VIG dataset, which suggests that the complexity inherent in driver drowsiness detection benefits profoundly from automated feature extraction. Furthermore, architectures specifically designed for EEG data, such as EEGNet, LMDA-Net, and TSception, generally surpass general-purpose models like the Transformer on the SEED-VIG task. This emphasizes the necessity of architectural choices that account for the non-stationary nature of neural signals. The success of the TSception framework suggests that parallel spatiotemporal pathways are highly effective for extracting diverse neurological markers. While the numerical improvement in accuracy for the proposed modified version is modest, the reduced CI indicates a vital enhancement in consistency. The state-of-the-art performance on the STEW dataset underscores the contribution of this study to multi-task cognitive monitoring, suggesting that a unified architecture can effectively serve both safety and productivity assessment domains.

When situating these results within the broader literature, the Modified TSception compares favorably against other specialized EEG architectures. In driver drowsiness detection, while hybrid models incorporating LSTM layers can achieve high accuracy, they often suffer from significant `black-box' complexity. The proposed model outperforms established benchmarks like EEGNet (80.74\%) and LMDA-Net (81.06\%) while prioritizing performance stability. The reduction of the CI to $\pm$0.24 represents a distinct improvement over the original TSception's $\pm$0.36. The significant performance drop of the Transformer model on SEED-VIG (68.95\%) compared to STEW (95.32\%) is likely attributable to its reliance on global attention, which struggles with the high noise floor and local transient patterns, such as theta bursts, characteristic of drowsiness data. In contrast, the multi-scale inception kernels of the Modified TSception filter these local signatures more effectively. The impact of ADP is further evidenced by the 95.93\% accuracy on the STEW dataset. While EEGNet (84.33\%) utilizes fixed-size kernels, the adaptive approach allows for the handling of specific signal lengths and channel configurations of consumer headsets without the loss of critical information during dimensionality reduction.

The reduction in performance variability is primarily attributed to the hierarchical refinement of temporal features. By capturing brain dynamics at multiple scales, the model is rendered more robust against the signal non-stationarity that frequently complicates EEG analysis. Several avenues for future research have been identified based on these findings. Future work will focus on the optimization of the Modified TSception for real-time processing on low-power edge devices to enable integration into commercial vehicle safety systems. Additionally, unsupervised domain adaptation techniques will be investigated to eliminate the requirement for subject-specific training. The integration of this architecture with other physiological signals, such as ocular or cardiac data, is expected to provide a more holistic view of cognitive states. Finally, the scalability of the ADP mechanism will be evaluated when transitioning to high-density EEG configurations, and the model's robustness will be tested in 'in-the-wild' scenarios to ensure performance against unpredictable real-world artifacts.

\textbf{Limitations and Constraints:} Despite the improved stability of the Modified TSception, certain constraints must be acknowledged. First, data quality varies between datasets, as SEED-VIG utilized research-grade equipment while STEW relied on a consumer-grade Emotiv EPOC headset. The latter is more susceptible to motion artifacts, necessitating the intensive ICA-based preprocessing employed in the pipeline. Second, demographics were limited to healthy young adults, and the neurological signatures of drowsiness in older populations or clinical groups may differ. Lastly, while ADP handles varying channel counts, the performance of the model in high-density environments remains to be established, as increased dimensionality may introduce computational bottlenecks.


\section{Conclusion}
\label{sec_conclusion}

This study investigated the application of DL for driver drowsiness detection and mental workload assessment using the SEED-VIG and STEW datasets. A Modified TSception model was proposed, incorporating hierarchical refinements to spatiotemporal feature extraction. Experimental results demonstrated a clear advantage of specialized EEG-focused designs over traditional ML and general-purpose Transformer models. While achieving a comparable accuracy of 83.46\% on the SEED-VIG dataset, the proposed model exhibited a significantly reduced CI of $\pm$0.24 compared to the $\pm$0.36 of the original architecture, suggesting a substantial improvement in performance stability. Furthermore, state-of-the-art results were achieved on the STEW dataset for both 2-class and 3-class tasks. These architectural modifications effectively improve model reliability, providing a robust framework for safety-critical BCIs. This study underscores the importance of tailored architectures for reliable cognitive monitoring and offers a promising path for future research in reducing the subject-variability gap for real-world deployment.




\section*{Bibliography}
\bibliography{references}

@Article{amer2023eeg,
  author    = {Amer, Nisreen Said and Belhaouari, Samir Brahim},
  journal   = {IEEE Access},
  title     = {{EEG Signal Processing for Medical Diagnosis, Healthcare, and Monitoring: A Comprehensive Review}},
  year      = {2023},
  pages     = {143116--143142},
  volume    = {11},
  doi       = {10.1109/access.2023.3341419},
  publisher = {IEEE},
}

@Article{balam2021statistical,
  author    = {Balam, Venkata Phanikrishna and Chinara, Suchismitha},
  journal   = {IEEE Transactions on Instrumentation and Measurement},
  title     = {{Statistical Channel Selection Method for Detecting Drowsiness through Single-Channel EEG-Based BCI System}},
  year      = {2021},
  pages     = {1--9},
  volume    = {70},
  doi       = {10.1109/tim.2021.3094619},
  publisher = {IEEE},
}

@Article{budak2019effective,
  author    = {Budak, Umit and Bajaj, Varun and Akbulut, Yaman and Atila, Orhan and Sengur, Abdulkadir},
  journal   = {IEEE Sensors Journal},
  title     = {{An Effective Hybrid Model for EEG-Based Drowsiness Detection}},
  year      = {2019},
  number    = {17},
  pages     = {7624--7631},
  volume    = {19},
  doi       = {10.1109/jsen.2019.2917850},
  publisher = {IEEE},
}

@Article{chaabene2021convolutional,
  author    = {Chaabene, Siwar and Bouaziz, Bassem and Boudaya, Amal and H{\"o}kelmann, Anita and Ammar, Achraf and Chaari, Lotfi},
  journal   = {Sensors},
  title     = {{Convolutional Neural Network for Drowsiness Detection Using EEG Signals}},
  year      = {2021},
  number    = {5},
  pages     = {1734},
  volume    = {21},
  doi       = {10.3390/s21051734},
  publisher = {MDPI},
}

@Article{chen2015automatic,
  author    = {Chen, Lan-lan and Zhao, Yu and Zhang, Jian and Zou, Jun-zhong},
  journal   = {Expert Systems with Applications},
  title     = {{Automatic Detection of Alertness/Drowsiness from Physiological Signals Using Wavelet-Based Nonlinear Features and Machine Learning}},
  year      = {2015},
  number    = {21},
  pages     = {7344--7355},
  volume    = {42},
  doi       = {10.1016/j.eswa.2015.05.028},
  publisher = {Elsevier},
}

@Article{chinara2021automatic,
  author    = {Chinara, Suchismitha and others},
  journal   = {Journal of Neuroscience Methods},
  title     = {{Automatic Classification Methods for Detecting Drowsiness Using Wavelet Packet Transform Extracted Time-Domain Features from Single-Channel EEG Signal}},
  year      = {2021},
  pages     = {108927},
  volume    = {347},
  doi       = {10.1016/j.jneumeth.2020.108927},
  publisher = {Elsevier},
}

@Article{cortes1995support,
  author    = {Cortes, Corinna and Vapnik, Vladimir},
  journal   = {Machine Learning},
  title     = {{Support-Vector Networks}},
  year      = {1995},
  pages     = {273--297},
  volume    = {20},
  doi       = {10.1007/bf00994018},
  publisher = {Springer},
}

@Article{craik2019deep,
  author    = {Craik, Alexander and He, Yongtian and Contreras-Vidal, Jose L},
  journal   = {Journal of Neural Engineering},
  title     = {{Deep Learning for Electroencephalogram (EEG) Classification Tasks: A Review}},
  year      = {2019},
  number    = {3},
  pages     = {031001},
  volume    = {16},
  doi       = {10.1088/1741-2552/ab0ab5},
  publisher = {IOP Publishing},
}

@Article{cui2022compact,
  author    = {Cui, Jian and Lan, Zirui and Liu, Yisi and Li, Ruilin and Li, Fan and Sourina, Olga and M{\"u}ller-Wittig, Wolfgang},
  journal   = {Methods},
  title     = {{A Compact and Interpretable Convolutional Neural Network for Cross-Subject Driver Drowsiness Detection from Single-Channel EEG}},
  year      = {2022},
  pages     = {173--184},
  volume    = {202},
  doi       = {10.1016/j.ymeth.2021.04.017},
  publisher = {Elsevier},
}

@Article{ding2022tsception,
  author    = {Ding, Yi and Robinson, Neethu and Zhang, Su and Zeng, Qiuhao and Guan, Cuntai},
  journal   = {IEEE Transactions on Affective Computing},
  title     = {{TSception: Capturing Temporal Dynamics and Spatial Asymmetry from EEG for Emotion Recognition}},
  year      = {2022},
  doi       = {10.1109/taffc.2022.3169001},
  publisher = {IEEE},
}

@Article{ding2024novel,
  author    = {Ding, Wenlong and Liu, Aiping and Guan, Ling and Chen, Xun},
  journal   = {IEEE Transactions on Neural Systems and Rehabilitation Engineering},
  title     = {{A Novel Data Augmentation Approach Using Mask Encoding for Deep Learning-Based Asynchronous SSVEP-BCI}},
  year      = {2024},
  pages     = {875--886},
  volume    = {32},
  doi       = {10.1109/tnsre.2024.3366930},
  publisher = {IEEE},
}

@Article{larocco2020systemic,
  author    = {LaRocco, John and Le, Minh Dong and Paeng, Dong-Guk},
  journal   = {Frontiers in Neuroinformatics},
  title     = {{A Systemic Review of Available Low-Cost EEG Headsets Used for Drowsiness Detection}},
  year      = {2020},
  pages     = {553352},
  volume    = {14},
  doi       = {10.3389/fninf.2020.553352},
  publisher = {Frontiers},
}

@Article{lawhern2018eegnet,
  author    = {Lawhern, Vernon J and Solon, Amelia J and Waytowich, Nicholas R and Gordon, Stephen M and Hung, Chou P and Lance, Brent J},
  journal   = {Journal of Neural Engineering},
  title     = {{EEGNet: A Compact Convolutional Neural Network for {EEG}-Based Brain--Computer Interfaces}},
  year      = {2018},
  number    = {5},
  pages     = {056013},
  volume    = {15},
  doi       = {10.1088/1741-2552/aace8c},
  publisher = {iOP Publishing},
}

@InProceedings{liu2022convnet,
  author    = {Liu, Zhuang and Mao, Hanzi and Wu, Chao-Yuan and Feichtenhofer, Christoph and Darrell, Trevor and Xie, Saining},
  booktitle = {Proceedings of the IEEE/CVF Conference on Computer Vision and Pattern Recognition},
  title     = {{A ConvNet for the 2020s}},
  year      = {2022},
  pages     = {11976--11986},
  doi       = {10.1109/cvpr52688.2022.01167},
}

@Article{miao2023lmda,
  author    = {Miao, Zhengqing and Zhao, Meirong and Zhang, Xin and Ming, Dong},
  journal   = {Neuroimage},
  title     = {{LMDA-Net: A Lightweight Multi-Dimensional Attention Network for General EEG-Based Brain-Computer Interfaces and Interpretability}},
  year      = {2023},
  pages     = {120209},
  volume    = {276},
  doi       = {10.1016/j.neuroimage.2023.120209},
  publisher = {Elsevier},
}

@Article{michel2019eeg,
  author    = {Michel, Christoph M and Brunet, Denis},
  journal   = {Frontiers in Neurology},
  title     = {{EEG Source Imaging: A Practical Review of the Analysis Steps}},
  year      = {2019},
  pages     = {325},
  volume    = {10},
  doi       = {10.3389/fneur.2019.00325},
  publisher = {Frontiers Media SA},
}

@Article{nguyen2017utilization,
  author    = {Nguyen, Thien and Ahn, Sangtae and Jang, Hyojung and Jun, Sung Chan and Kim, Jae Gwan},
  journal   = {Scientific Reports},
  title     = {{Utilization of a Combined EEG/NIRS System to Predict Driver Drowsiness}},
  year      = {2017},
  number    = {1},
  pages     = {43933},
  volume    = {7},
  doi       = {10.1038/srep43933},
  publisher = {Nature Publishing Group UK London},
}

@Article{ogino2018portable,
  author    = {Ogino, Mikito and Mitsukura, Yasue},
  journal   = {Sensors},
  title     = {{Portable Drowsiness Detection through Use of a Prefrontal Single-Channel Electroencephalogram}},
  year      = {2018},
  number    = {12},
  pages     = {4477},
  volume    = {18},
  doi       = {10.3390/s18124477},
  publisher = {MDPI},
}

@Article{paulo2021cross,
  author    = {Paulo, Joao Ruivo and Pires, Gabriel and Nunes, Urbano J},
  journal   = {IEEE Transactions on Neural Systems and Rehabilitation Engineering},
  title     = {{Cross-Subject Zero Calibration Driver's Drowsiness Detection: Exploring Spatiotemporal Image Encoding of EEG Signals for Convolutional Neural Network Classification}},
  year      = {2021},
  pages     = {905--915},
  volume    = {29},
  doi       = {10.1109/tnsre.2021.3079505},
  publisher = {IEEE},
}

@Article{perkins2022challenges,
  author    = {Perkins, Emma and Sitaula, Chiranjibi and Burke, Michael and Marzbanrad, Faezeh},
  journal   = {IEEE Transactions on Intelligent Vehicles},
  title     = {{Challenges of Driver Drowsiness Prediction: The Remaining Steps to Implementation}},
  year      = {2022},
  number    = {2},
  pages     = {1319--1338},
  volume    = {8},
  doi       = {10.1109/tiv.2022.3224690},
  publisher = {IEEE},
}

@Article{roy2019deep,
  author    = {Roy, Yannick and Banville, Hubert and Albuquerque, Isabela and Gramfort, Alexandre and Falk, Tiago H and Faubert, Jocelyn},
  journal   = {Journal of Neural Engineering},
  title     = {{Deep Learning-Based Electroencephalography Analysis: A Systematic Review}},
  year      = {2019},
  number    = {5},
  pages     = {051001},
  volume    = {16},
  doi       = {10.1088/1741-2552/ab260c},
  publisher = {IOP Publishing},
}

@Article{samiee2014data,
  author    = {Samiee, Sajjad and Azadi, Shahram and Kazemi, Reza and Nahvi, Ali and Eichberger, Arno},
  journal   = {Sensors},
  title     = {{Data Fusion to Develop a Driver Drowsiness Detection System with Robustness to Signal Loss}},
  year      = {2014},
  number    = {9},
  pages     = {17832--17847},
  volume    = {14},
  doi       = {10.3390/s140917832},
  publisher = {Multidisciplinary Digital Publishing Institute},
}

@Article{siddhad2024efficacy,
  author    = {Siddhad, Gourav and Gupta, Anmol and Dogra, Debi Prosad and Roy, Partha Pratim},
  journal   = {Biomedical Signal Processing and Control},
  title     = {{Efficacy of Transformer Networks for Classification of EEG Data}},
  year      = {2024},
  pages     = {105488},
  volume    = {87},
  doi       = {10.1016/j.bspc.2023.105488},
  eprint    = {2202.05170},
  publisher = {Elsevier},
}

@Article{siddhad2024enhancing,
  author   = {Siddhad, Gourav and Iwamura, Masakazu and Roy, Partha Pratim},
  journal  = {IEEE Transactions on Artificial Intelligence},
  title    = {{Enhancing EEG Signal-Based Emotion Recognition with Synthetic Data: Diffusion Model Approach}},
  year     = {2025},
  pages    = {1-10},
  doi      = {10.1109/TAI.2025.3641576},
  eprint   = {2401.16878},
}

@Article{sklearn,
  author  = {Pedregosa, F. and Varoquaux, G. and Gramfort, A. and Michel, V. and Thirion, B. and Grisel, O. and Blondel, M. and Prettenhofer, P. and Weiss, R. and Dubourg, V. and Vanderplas, J. and Passos, A. and Cournapeau, D. and Brucher, M. and Perrot, M. and Duchesnay, E.},
  journal = {Journal of Machine Learning Research},
  title   = {{Scikit-Learn: Machine Learning in Python}},
  year    = {2011},
  pages   = {2825--2830},
  volume  = {12},
}

@Article{stancin2021review,
  author    = {Stancin, Igor and Cifrek, Mario and Jovic, Alan},
  journal   = {Sensors},
  title     = {{A Review of EEG Signal Features and Their Application in Driver Drowsiness Detection Systems}},
  year      = {2021},
  number    = {11},
  pages     = {3786},
  volume    = {21},
  doi       = {10.3390/s21113786},
  publisher = {MDPI},
}

@Article{tarafder2022drowsiness,
  author    = {Tarafder, Sreeza and Badruddin, Nasreen and Yahya, Norashikin and Nasution, Arbi Haza},
  journal   = {Sensors},
  title     = {{Drowsiness Detection Using Ocular Indices from EEG Signal}},
  year      = {2022},
  number    = {13},
  pages     = {4764},
  volume    = {22},
  doi       = {10.3390/s22134764},
  publisher = {MDPI},
}

@Article{turkoglu2021deep,
  author    = {Turkoglu, Muammer and Alcin, Omer F and Aslan, Muzaffer and Al-Zebari, Adel and Sengur, Abdulkadir},
  journal   = {Biomedical Signal Processing and Control},
  title     = {{Deep Rhythm and Long Short Term Memory-Based Drowsiness Detection}},
  year      = {2021},
  pages     = {102364},
  volume    = {65},
  doi       = {10.1016/j.bspc.2020.102364},
  publisher = {Elsevier},
}

@Article{walther2023systematic,
  author    = {Walther, Dominik and Viehweg, Johannes and Haueisen, Jens and M{\"a}der, Patrick},
  journal   = {Frontiers in Neuroinformatics},
  title     = {{A Systematic Comparison of Deep Learning Methods for EEG Time Series Analysis}},
  year      = {2023},
  pages     = {1067095},
  volume    = {17},
  doi       = {10.3389/fninf.2023.1067095},
  publisher = {Frontiers Media SA},
}

@Article{wei2018subject,
  author    = {Wei, Chun-Shu and Lin, Yuan-Pin and Wang, Yu-Te and Lin, Chin-Teng and Jung, Tzyy-Ping},
  journal   = {Neuroimage},
  title     = {{A Subject-Transfer Framework for Obviating Inter-and Intra-Subject Variability in EEG-Based Drowsiness Detection}},
  year      = {2018},
  pages     = {407--419},
  volume    = {174},
  doi       = {10.1016/j.neuroimage.2018.03.032},
  publisher = {Elsevier},
}

@Article{zheng2017multimodal,
  author    = {Zheng, Wei-Long and Lu, Bao-Liang},
  journal   = {Journal of Neural Engineering},
  title     = {{A Multimodal Approach to Estimating Vigilance Using EEG and Forehead EOG}},
  year      = {2017},
  number    = {2},
  pages     = {026017},
  volume    = {14},
  doi       = {10.1088/1741-2552/aa5a98},
  publisher = {iOP Publishing},
}

@Article{lim2018stew,
  author    = {Lim, Wei Lun and Sourina, Olga and Wang, Lipo P},
  journal   = {IEEE Transactions on Neural Systems and Rehabilitation Engineering},
  title     = {{STEW: Simultaneous Task EEG Workload Data Set}},
  year      = {2018},
  number    = {11},
  pages     = {2106--2114},
  volume    = {26},
  doi       = {10.1109/tnsre.2018.2872924},
  publisher = {IEEE},
}








\end{document}